\documentclass[aps,pra,reprint,superscriptaddress]{revtex4-1}

\usepackage[utf8]{inputenc} 
\usepackage{amsmath} 
\usepackage{amssymb} 
\usepackage{graphicx} 
\usepackage{braket} 
\usepackage[textsize=tiny]{todonotes}

\newcommand{\dee}[0]{\mathrm{d}}

\DeclareGraphicsExtensions{.eps}

\begin{document}

\title{Modeling electron dynamics coupled to continuum states in finite volumes}

\author{Umberto De Giovannini} \email[]{umberto.degiovannini@ehu.es}

\affiliation{Nano-Bio Spectroscopy Group and ETSF Scientific Development Centre, Departamento de Física de Materiales, Universidad del País Vasco, CSIC-UPV/EHU-MPC and DIPC, Avenida de Tolosa 72, E-20018 San Sebastián, Spain}

\author{Ask Hjorth Larsen} 
\affiliation{Nano-Bio Spectroscopy Group and ETSF Scientific Development Centre, Departamento de Física de Materiales, Universidad del País Vasco, CSIC-UPV/EHU-MPC and DIPC, Avenida de Tolosa 72, E-20018 San Sebastián, Spain}

\author{Angel Rubio} 
\email{angel.rubio@ehu.es}
\affiliation{Nano-Bio Spectroscopy Group and ETSF Scientific Development Centre, Departamento de Física de Materiales, Universidad del País Vasco, CSIC-UPV/EHU-MPC and DIPC, Avenida de Tolosa 72, E-20018 San Sebastián, Spain}
\affiliation{Fritz-Haber-Institut der Max-Planck-Gesellschaft, Berlin, Germany}



\date{\today}
\begin{abstract}
  Absorbing boundaries are frequently employed in real-time propagation of the Schr\"odinger equation to remove
  spurious reflections and efficiently emulate outgoing boundary conditions. 
  These conditions are a fundamental ingredient for 
  an implicit description of observables involving infinitely extended continuum states. 
  In the literature, several boundary absorbers have been proposed. They mostly fall into three main families: 
  mask function absorbers, complex absorbing potentials, and exterior complex-scaled potentials.
  To date none of the proposed absorbers is perfect, and all present a certain degree of reflections. 
  Characterization of such reflections is thus a critical task with strong implications for time-dependent 
  simulations of atoms and molecules.
  We introduce a method to evaluate the reflection properties of a given absorber and present a comparison 
  of selected samples for each family of absorbers.  
  Further, we discuss the connections between members of each family and show how the same reflection 
  curves can be obtained with very different absorption schemes.
\end{abstract}

\maketitle

\section*{Introduction} 
\label{sec:intro} 
Continuum states play a central role in many physical processes
involving atoms, molecules and extended low-dimensional systems.
They are fundamental to the description of scattering problem with several applications 
ranging from ion collision and molecular damage~\cite{Boudaffa:2000ji} to electronic transport in molecular 
junctions~\cite{Cuniberti:2006dk}.
Besides being important as final states there are situations where they play more articulated roles.
This is the case, for instance, in strong laser physics where,
in addition to direct ionization which involves the electronic continuum as final state, continuum states act
as intermediate configurations before rescattering giving origin to peculiar features in 
photoemission~\cite{Meckel:2008vn,Huismans:2011kh} and high harmonic generation~\cite{Lock:2012wa}. 

In discrete and finite-size representations, either in real (e.g. finite differences) or Hilbert (e.g. basis set discetization) spaces, 
a detailed theoretical description of continuum states is a difficult task. 
The main difficulty arises from their extended nature spanning the whole space.
In molecules the situation is further complicated by the presence of non-spherical multi-center potentials
which quickly render any analytic approach impossible as the system grows in size.
Several methods have been employed to address the problem; 
among the others, we mention the iterative
Schwinger method~\cite{Lucchese:1982fga}, and B-spline expansions~\cite{Sanchez:1997jo,Bachau:2001gm}.

In many situations an explicit representation of the continuum wavefunctions 
is not necessary to calculate the relevant observables involving unbound states.
Problems such as the description of ionization processes induced by external fields, 
the calculation of the absorption cross-section of atoms and molecules in the continuum, or 
electronic transport through molecular junctions, are just few prototypical examples.
In these cases it is convenient to partition the space into two regions: 
an inner region containing the atomic or molecular system under study 
and an outer region defined as its complement.
In the inner region the Schr\"odinger equation can be numerically solved with computational tools appropriate
for bound electronic configurations.
Owing to the absence of the nuclear centers, the Hamiltonian in the outer region is greatly simplified and can  
often be treated semi-analytically. 
The continuum can then be accounted for by matching inner and outer solutions.
The problem is thus recast from finding an explicit representation of the continuum states to an implicit one
where one solves the inner problem,
imposing the appropriate conditions at the boundary between the two regions.
Techniques used in these cases span from the R-matrix 
theory~\cite{ Dora:2009ef,Descouvemont:2010do} to Green's function methods~\cite{Ermolaev:1999bp} which have been 
applied both to the free~\cite{Inglesfield:2008br,Inglesfield:2011hs,Nakatsukasa:2001hv} (e.g.~for ionization processes) and bulk-continuum cases~\cite{Kurth:2005bc,Stefanucci:2008bp} (e.g.~for electron transport).

Absorbing boundaries (ABs) or boundary absorbers fall into the category of techniques employed in the former
situation, i.e., when explicit description of continuum states is not needed.
Their simplicity of implementation and limited computational cost make
them, in many situations, preferable to the previously mentioned approaches.
The idea is to insert a buffer region of a given width at the edges of the simulation box, where 
the absorber acts by removing unwanted reflections.
The extent to which reflections are removed in a given energy range is in turn
related to how well continuum states within the same energy range are described.

In this work we focus on three families of absorbers: mask function absorbers (MFAs), complex absorbing 
potentials (CAPs), and absorbers derived from smooth-exterior complex scaling (SES).

Mask function absorbers are the simplest type of absorber.  
At each time step one multiplies the wavefunction by a mask function,
which has values between zero and one depending on the region in space.  The function is generally chosen
to be 1 within an inner region, then decay to zero near the boundary of the simulation box so as to
damp any part of the wavefunction extending towards the boundary.
MFAs have been employed in many different situations.
We mention strong laser field studies on high harmonic generation~\cite{Krause:1992dea}, electron and proton 
emission~\cite{Kulander:1996go}, and above-threshold ionization~\cite{Lein:2002ft}.
They are the cornerstone of split-domain propagation schemes~\cite{Chelkowski:1998ec,Grobe:1999vy} where they 
act as matching layers between two different domains.
In the literature they have been mainly associated with few-body problems (two or three particles),  
but in recent works they have also been used to study 
electron photoemission in multi-electron systems combined with time-dependent density functional 
theory~\cite{UDeGiovannini:2012hy,DeGiovannini:2013dr, CrawfordUranga:2014bd}.

The complex absorbing potential is, by far, the most popular type of absorber.
The method consists in altering the Hamiltonian of a system by adding an artificial 
complex potential which is non-zero in a small portion of the simulation box close to the boundaries.
Time propagation with this modified Hamiltonian results in wavepacket absorption or injection from the boundaries.
Following their introduction as negative imaginary potentials in the work of Neuhauser and 
Baer~\cite{Neuhauser:1989fa}, 
CAPs have been applied to a wealth of different situations ranging from relaxation of holes states 
in clusters~\cite{Santra:2002gy} to charge injection for transport problems~\cite{Varga:2007ia, Wibking:2012bs}.
We refer the reader to an excellent review on the subject 
for further information~\cite{MUGA:2004jy}.

The complex scaling method was originally developed to calculate the
properties of metastable states or resonances.  The method has a solid
theoretical background with a long history of development, beginning
with the original theorems by Aguilar, Balslev and
Combes~\cite{aguilarcombes,Balslev:1971ez}.
The method continue to be the subject of much interest
and have since been adapted to scalable computational approaches including
configuration interaction~\cite{PhysRevA.66.052713} and DFT~\cite{Whitenack:2011ga,Larsen:2013cw}.
Complex scaling methods work by
means of a scaling of the position variable of the Hamiltonian
by a complex factor.
This causes states that represent extended out-going waves or resonances to become normalizable and
thus computationally accessible.\cite{simon1973resonances}

Exterior complex scaling (ECS), proposed by Simon~\cite{Simon:1979tc},
is a generalization where the scaling transformation is only applied
outside a certain region rather than uniformly.  
ECS was originally introduced
to avoid singularities in the Coulomb interaction at nuclear centers where the scaling method breaks down. 
In smooth-exterior scaling (SES) the
scaling is governed by a smooth function to improve numerics.

Complex scaling methods can be applied to time-dependent problems as well as static ones.
Under time-propagation, states extending into the complex-scaled (exterior) region will decay over
time.  The region thus acts as an absorber~\cite{McCurdy:1991jp,Riss:1999ez,Moiseyev:1999hs}.
The method is also referred to as a reflection-free CAP (RF-CAP) even though it was shown to present reflections 
in combination with finite difference implementations~\cite{Shemer:2005jh}.
It was later shown by Scrinzi~\cite{Scrinzi:2010jv} that such reflections can be reduced to 
machine precision in time propagations under strong infrared fields using a finite element approach.

As far as finite difference methods are concerned, none of the ABs known in the literature is numerically 
reflection-free and, depending on the type of AB and the size of the buffer region, their absorption properties vary 
as a function of the impinging wavefunction's kinetic energy. 
Characterization and control of the reflection properties of an AB is thus an important task with implications
to many situations involving an implicit description of continuum states.
By employing tailored Gaussian wavepackets we are able to map the reflection properties of a given absorber
and display it as a function of different parameters. 
In the present work, we use these properties to illustrate differences and analogies between some of 
the most used absorbers.

\begin{figure}[ht]
	\includegraphics[width=\columnwidth]{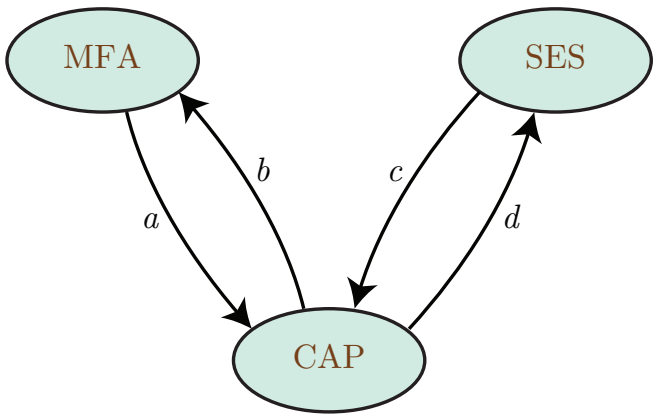} 
	\caption{The connections between the three different families of ABs covered by the present work.
   \label{fig:scheme}} 
\end{figure}

In some cases different types of AB behave very similarly, indicating a degree of 
relation.
In fact, it can be shown that all the three families are connected to one another as in Fig.~\ref{fig:scheme}. 
While each of the single connections in the figure has been separately discussed in the literature, 
in this paper we aim at presenting the full picture in an homogeneous framework. 
Furthermore, besides illustrating the nature of each link, we show how it is possible to obtain the 
same absorption properties with very different types of ABs.

The paper is organized as follows. We summarize the scattering theory from complex potentials in 
Sec.~\ref{sec:theory} and introduce the quantities used to assess the reflection properties 
of ABs in finite volumes. 
In Sec.~\ref{sec:results} we describe the details of the numerical calculations.
We discuss the reflection properties of MFAs in Sec.~\ref{sec:mask_function_absorbing_boundaries}. 
In Sec.~\ref{sec:cap} we present the reflection properties of polynomial and $\sin^2$ CAPs for different 
potential heights and illustrate by calculating the absorption cross-section of a one-dimensional hydrogen 
atom.  
The connection between MFAs and CAPs is discussed in Sec.~\ref{sec:mask_and_cap}.
Finally in Sec.~\ref{sec:ses} we discuss SES in connection with CAPs.

\section{Theory} 

\label{sec:theory} 
Owing to the position they cover in Fig.~\ref{fig:scheme}, CAPs play a central 
role in the present work.
Below we briefly review the scattering theory from complex potentials following the review by 
Muga.~\cite{MUGA:2004jy}. 
Besides introducing the reader to the main theoretical formalism the purpose of the present 
section is that of formally derive $\epsilon(k_0)$, a quantity assessing the error committed by an AB 
that is suitable for numerical evaluation in finite volumes.
Atomic units will be used throughout ($m_{e}=e=\hbar= 1$).

\begin{figure}[ht]
  \includegraphics[width=\columnwidth]{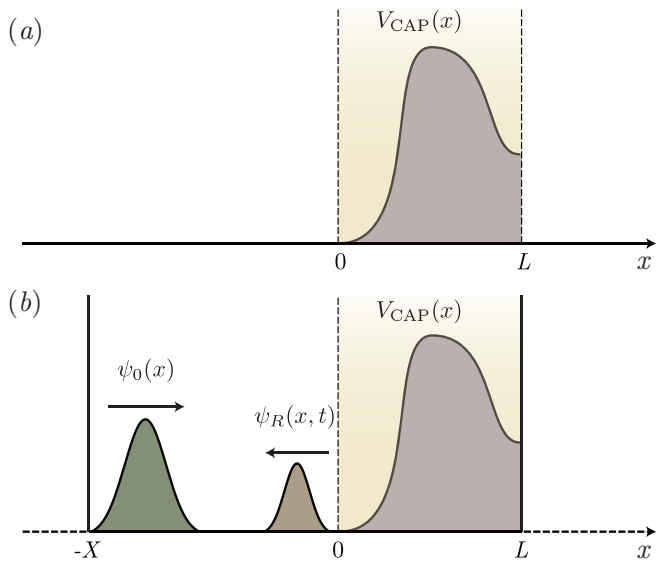}	
  \caption{Scattering from a CAP in (a) infinite and (b) finite volumes.
  \label{fig:scattering}} 
\end{figure}

In one-dimension, the stationary Schr\"odinger equation for a free electron in the presence 
of a CAP may be written as 
\begin{equation}\label{eq:capH}
	H\psi(x) = H_0 \psi(x) + V_{\rm CAP}(x)\psi(x)= E\psi(x) 
\end{equation}
with $H_0$ being the physical Hamiltonian and $V_{\rm CAP}(x)$ an absorbing potential with finite support in the 
region $[0,L]$ as in Fig.~\ref{fig:scattering}~(a).
For simplicity we consider here the case where $H_0$ describes free electrons $H_0=- 
\nabla^2/2$. 

The non-normalizable solutions of \eqref{eq:capH} corresponding to stationary scattering states are obtained 
by imposing one of the following asymptotic boundary conditions, 
\begin{equation}
	\label{eq:asym_left} \psi^{+}(k,x)=\left\{ 
	\begin{array}{ll}
		e^{i k x} + R^l(k)e^{-i k x} \qquad & x \rightarrow -\infty\\
		T^l(k) e^{i k x} \qquad & x\rightarrow \infty 
	\end{array}
	\right. 
\end{equation}
\begin{equation}
	\label{eq:asym_right} \psi^{-}(k,x)=\left\{ 
	\begin{array}{ll}
		T^r(k) e^{-i k x} \qquad & x \rightarrow -\infty\\
		e^{-i k x} + R^r(k)e^{+i k x}\qquad & x\rightarrow \infty 
	\end{array}
	\right. \,. 
\end{equation}
For $k>0$, they enforce the condition corresponding respectively to a right-travelling wave $\Psi^+(k,x)$ 
and a left-travelling wave $\Psi^{-}(k,x)$.
The functions $R^{l,r}(k)$ and $T^{l,r}(k)$ indicate the (left/right) reflection and transmission 
coefficients. 
These coefficients are, in general, complex functions of $k$ and their square modulus represents the probability 
to reflect and transmit a wave with a given momentum.
If the initial wavepacket is localized on the left side of the potential 
we can simplify the notation by imposing only the left asymptotic condition,
discarding the right quantities.

By imposing the left asymptotic boundary conditions of~\eqref{eq:asym_left} on the time-dependent Schr\"odinger
equation   
\begin{equation}
   i\frac{\partial \psi(t,x)}{\partial t} = H\psi(t,x)\,,
\end{equation}
a unique solution is obtained for every single initial condition.
For a right-moving wavepacket $\psi_0(x)$ 
initially localized to the left region $[-\infty, 0]$, this solution is given by
\begin{equation}
	\label{eq:psit_infinite} 
  \psi(t,x)=\int_{0}^{\infty} {\rm d} k \, \psi^{+}(k,x) e^{- i k^2 t/2 }\psi_0(k)\,, 
\end{equation}
where space is restricted to $x \in [-\infty, 0] \cup [L, \infty]$, and $\psi_0(k)$ represents the Fourier transform of $\psi_0(x)$. 
Here, $U(t)=e^{- i k^2 t/2}$ is the time evolution operator associated with the 
free Hamiltonian $H_0$. 
Note that, although reflection and absorption coefficients are strictly defined for $k>0$, they can be analytically continued  
to negative $k$ and the integral \eqref{eq:psit_infinite} can be extended to the whole real axis. 

The most appropriate quantity to assess the absorption properties of a CAP is the survival probability $S(k)$. 
It is defined as the the sum of transmission $|T(k)|^2$ and reflection $|R(k)|^2$ probabilities:
$S(k) := |T(k)|^2 + |R(k)|^2$.
According to this definition, an ideal absorber, preventing reflections regardless of the 
incoming wavevector, would have $S(k)=0$ for all $k>0$. 
As already mentioned, none of the known CAPs is free from reflections, and the evaluation of $S(k)$
is of great importance for practical applications.

While from \eqref{eq:psit_infinite} it is possible to evaluate the time evolution of an 
arbitrary wavepacket knowing the reflection and transmission coefficients (and thus $S(k)$), 
the calculation of these coefficients starting from the knowledge of $\psi(t,x)$ is a more complicated task.
A widely used approach consists in calculating $R(k)$ and $T(k)$ by numerically solving the static one-dimensional 
scattering problem~\cite{Macias:1994we, Vibok:2001fp, Manolopoulos:2002kw}. 
This solution is however difficult to apply in the present work where we aim at 
addressing ABs expressed with quite different formulations.
We instead use a different approach based on direct time propagation in a finite volume.

We focus on the situation where the solution of \eqref{eq:capH} is confined to the volume $[-X,L]$
like in Fig.~\ref{fig:scattering}~(b). 
In this case it is customary to impose zero boundary conditions at the border of the region: i.e. $\psi(t,x=-X)=\psi(t,x=L)=0$ for all $t$. 
The CAP therefore effectively includes an infinite barrier at $x=L$. 
CAPs so defined are also known as Type I potentials~\cite{Palao:1998bv, MUGA:2004jy}. 
A wavefunction is thus either absorbed or reflected,  
the transmission coefficient $T(k)$ is therefore equal to zero and the survival probability is just equal to the square modulus of the reflection coefficient: $S(k)= |R(k)|^2$.  

The situation is similar to what illustrated in Fig.~\ref{fig:scattering}~(b): 
a left-moving wavepacket $\psi_0(x)$ initially localized in $[-X,0]$ propagates as the sum of a
free propagating and a reflected wave   
\begin{eqnarray}
	\label{eq:psit_finite} 
  \psi(t,x) &=&  \psi_0(t,x)+\psi_R(t,x)  \nonumber \\
  &=&\int_{-\infty}^{\infty} \frac{{\rm d} k}{\sqrt{2\pi}} \,e^{i k x} e^{- i k^2 t/2 }\psi_0(k) + \nonumber \\
	&&\int_{-\infty}^{\infty} \frac{{\rm d} k}{\sqrt{2\pi}} \,e^{-i k x} R(k) e^{- i k^2 t/2 }\psi_0(k)\,.
\end{eqnarray}
To obtain \eqref{eq:psit_finite} we extended the integral in \eqref{eq:psit_infinite} over negative $k$ and
imposed the normalization on incoming and outgoing waves.

All the informations regarding the reflection coefficient, and therefore $S(k)$, are contained in $\psi_R(t,x)$.
The reflection error~\cite{Zavin:1998dt}, defined as  
\begin{eqnarray}
  \label{eq:epsilontilde}
  \tilde{\epsilon}(t) &=& \int_{-X}^{0}{\rm d} x\, |\psi(t,x) - \psi_{0}(t,x)|^2\nonumber \\
  &=& \int_{-X}^{0}{\rm d} x\,|\psi_{R}(t,x)|^2\,,
\end{eqnarray} 
is the closest quantity related to $\psi_R(t,x)$ that can be easily accessed from the direct time evolution
of $\psi(t,x)$. 
In the long time limit $t\geq\tau$, when the free propagating wavefunction has left the 
simulation region, it represents a direct measure of reflections.
Provided that the initial wavepacket $\psi_0(k)$ is well localized around $k_0$, it directly approximates $S(k_0)$
and, in the limit where $\psi_0(k)$ is a Dirac delta function centered at $k_0$ and $X\rightarrow \infty$, it
conicindes with the survival probability: 
$\tilde{\epsilon}(t\geq\tau)= S(k_0)$ .

The time evolution described by \eqref{eq:psit_finite} is defined only outside the CAP region, i.e. in $[-X,0]$,
hence the integration range in the definition of $\tilde{\epsilon}(t)$ in \eqref{eq:epsilontilde}. 
For times bigger than $\tau$ such that $\psi_{0}(t=\tau,x)$ is mostly localized outside this region and 
$\psi_{R}(t=\tau,x)$ is mostly localized in $[-X,0]$, the integral can be extended to the whole range $[-X,L]$.
We can thus extend the definition of \eqref{eq:epsilontilde} to the entire simulation box including the 
absorbing boundary region.
For a given initial wavepacket $\psi_0(x)$ localized in $[-X,0]$ with a peak around $k_0$, the reflection 
error can be redefined as
\begin{equation}
  \label{eq:epsilon}
  \epsilon(k_0) = \int_{-X}^{L}{\rm d} x\,|\psi(t=\tau,x) - \psi_{0}(t=\tau,x)|^2\,,
\end{equation}
where $\tau$ must be chosen big enough to satisfy the localization conditions on $\psi_{0}(t,x)$ and
$\psi_{R}(t,x)$, but smaller than the time needed for $\psi_{R}(t,x)$ to bounce at $x=-X$ back into the 
absorption region.

This extended definition is preferable to that of \eqref{eq:epsilontilde} as it correctly handles 
pathological situations such as the one occurring when the CAP is just a perfect reflecting wall placed at $x=L$. 
In this case a normalized wavepacket would just bounce back at $x=L$ and reach $[-X,0]$ after a time that 
depends on $k_0$ and on the width of the absorber region $L$. 
The unmodified reflection error $\tilde{\epsilon(t)}$ would than report either high or negligible values 
depending on such time while $\epsilon(k_0)$ would be a constant strictly equal to 1 hence indicating 
maximum reflection.


\section{Results} 

\label{sec:results}

In order to evaluate the reflection error for a range of energies we need to specify a family
of initial wavefunctions $\psi_0(x).$
The Gaussian wavepackets
\begin{equation}
	\label{eq:gaussianwp} \mathcal{G}(x, x_0, k_0,\sigma) = \pi^{-\frac{1}{4}} \sigma^{-\frac{1}{2}} e^{-\frac{(x-x_0)^2}{2 \sigma^2}+i k_0 x} 
\end{equation}
are a natural choice, being localized in both real and Fourier space.
Similar choices have been made in the literature~\cite{Vibok:1992uh,Riss:1996iu,Zavin:1998dt}. 
Here \eqref{eq:gaussianwp} describes a normalized Gaussian wavepacket with initial velocity 
$k_0$, width $\sigma$, centered in $x_0$.  We specifically choose the parameters
\begin{align}
  \sigma&=4\sqrt2 / k_0\,,\\
  x_0&=-3\sigma=-12\sqrt2 / k_0\,.
\end{align}
Given an absorber width $L$, we then sample the reflection error $\epsilon(k_0,L)$ by performing time evolutions 
of Gaussian wavepackets with different $k_0$.  

In each propagation $\mathcal{G}(x, x_0, k_0,\sigma)$ is centered in $x_0=-3\sigma$ and is evolved for 
a total time $\tau= 2(3\sigma+L)/k_0$ in a simulation box with $X=6\sigma$. 
The total propagation time is chosen so that the free wavepacket travels a distance of $d=6\sigma + 2L$.
During the time evolution, $\mathcal{G}(x,x_0, k_0,\sigma)$ spreads with an asymptotic velocity that depends 
on $\sigma$: $v_{\sigma}=1/\sqrt{2}\sigma$. 
The wavepacket is thus characterized by the two velocities $k_0$ and $v_{\sigma}$.
The choice $\sigma(k_0)=4 \sqrt{2}/k_0$ therefore enforces 
an expansion velocity slower than the translational one: $v_{\sigma}=k_0/8$. 
The chosen parameters also impose negligible left-moving ($k<0$) components and ensure that
the wavepackets have well-defined energies
also when the energy is close to 0.
With these choices the only free parameters left are $k_0$, $L$.

All the numerical calculations have been performed using finite differences in real space and real time with the  
Octopus~\cite{OCTOPUS,OCTOPUS1, OCTOPUS2} and QuantumPy~\cite{quantumpy} codes.
The spatial coordinates are discretized on a grid with spacing 
$\Delta x = 0.1$~a.u.\ and the time step $\Delta t= 0.01$~a.u.\ ($\Delta x = 0.05$~a.u.\ and time step $\Delta 
t= 0.001$~a.u.\ for kinetic energies greater than 2~KeV). 
These parameters secure an accurate description of the time propagation for all the wavepackets
considered. 
Additionally, they are directly comparable to the typical ones used in three-dimensional atomic and 
molecular calculations. 
To take a step further in the direction of application to real systems, in what follows, we 
express the reflection error $\epsilon(E,L)$ as a function of the wavepacket kinetic energy $E= 5 k_0^2 /4$ (in eV)
and absorber width $L$ (in atomic units).

\section{Mask function absorbers (MFAs)} 
\label{sec:mask_function_absorbing_boundaries}
MFAs~\cite{Krause:1992dea} are commonly used in numerical propagation schemes where 
the infinitesimal time-evolution operator $U(t+\Delta t, t)$, connecting $t$ to $t+\Delta t$, is repeatedly applied
to an initial wavefunction.
One chooses a mask function $0\leq M(x)\leq1$ and apply it multiplicatively on each time step, i.e.
\begin{equation}\label{eq:MaskProp}
  \psi(x,t+\Delta t) = M(x)U(t+\Delta t, t)\psi(x,t)\,. 
\end{equation}
If the function $M(x)$ is chosen to smoothly decay from one in the buffer region $[0,L]$,
the iterative application of \eqref{eq:MaskProp} results in a damping of
any part of the wave extending into this region.
In order to absorb, the mask function has to be smaller than 1 but is free to assume any value 
from 0 to 1 at the border $x=L$. 
The resulting absorption properties depend on the specific functional form of $M(x)$.
\begin{figure}[ht]
	\includegraphics[width=\columnwidth]{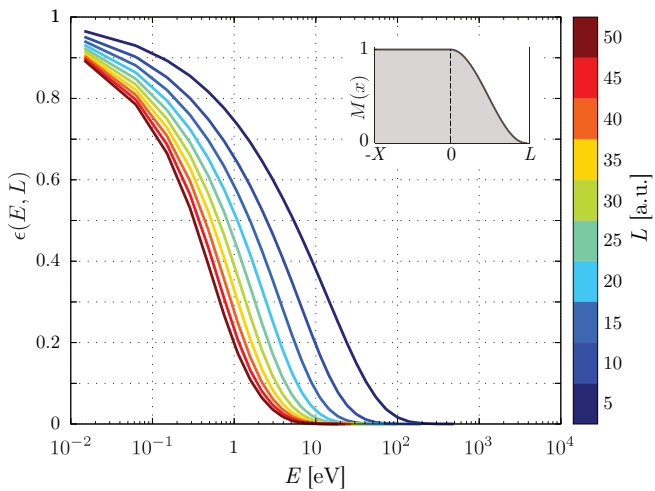} 
	\caption{Reflection error $\epsilon(E,L)$ as a function of kinetic energy $E$ (in logarithmic scale) and boundary width $L$, for the mask function absorber with $M(x)$ as defined by \eqref{eq:Mask_sin2} and shown in the inset.
   \label{Mask_sin2:fig}} 
\end{figure}

In Fig.~\ref{Mask_sin2:fig} we show the reflection error $\epsilon(E,L)$ as function of  
kinetic energy $E$ and boundary width $L$ for a mask function defined as
\begin{equation}
	\label{eq:Mask_sin2} M(x)= \left\{ 
	\begin{array}{ll}
		1 & \mbox{if $x < 0$} \\
		1-\sin^{2}\left[ \frac{ x \pi}{2L} \right] & \mbox{if $0\leq x \leq L $} 
	\end{array}
	\right. \,. 
\end{equation}
Reflections smoothly decay from almost unity at low energies to negligible values for high energies.
The range of energies with minimal reflection is connected to the boundary width $L$: the
greater $L$, the wider the region of minimal reflection. 
The high reflectivity for low energies can be diminished by 
increasing $L$, but slowly moving wavepackets, associated with large wavelengths, are difficult to remove with 
finite size absorbers. 
In the limit $L\rightarrow \infty$, the reflection is not going to zero for all
$E$ and $\epsilon(E,L)$ appears converging to a shape that depends on the functional form of $M(x)$.
As we will show, this is a common feature shared among all the ABs treated in the present work.

Besides being easy to implement, the main attraction of MFAs derives from the central 
role they play in split-domain propagation
schemes~\cite{Chelkowski:1998ec,Grobe:1999vy}. 
At the foundation of these schemes lies the idea that it is always possible to 
perform a spatial partitioning of a wavefunction as a sum of two partially overlapping components 
\begin{equation}
  \label{eq:masksplit}
  \psi(x,t) = M(x)\psi(x,t) +(1-M(x)) \psi(x,t) \,.
\end{equation}
In this partitioning, one piece of the wavefunction is localized in $[-\infty,L]$ and the other in 
the partially overlapping region $[0,\infty]$. 
The two parts can then be separately propagated with different Hamiltonians using different approximations. 
This decomposition is at the core of a recent method developed for the 
calculation of electron photoemission in molecular systems~\cite{UDeGiovannini:2012hy}.

\section{Complex absorbing potentials (CAPs)} 

\label{sec:cap} 
ABs based on CAPs are constructed by adding an artificial complex potential $V_{\rm CAP}(x)$ to the physical 
Hamiltonian $H_0$ as described by \eqref{eq:capH}.
By virtue of this change, the new Hamiltonian $H$ is no longer Hermitian, and  
the associated time propagation operator, 
\begin{equation}
  \label{eq:ucap}
  U_{CAP}(t+\Delta t, t)=e^{-i H \Delta t}=e^{-i [H_0+V_{\rm CAP}(x)] \Delta t} \,,
\end{equation}
becomes non-unitary. 
When $V_{\rm CAP}(x)$ is chosen to be non-zero only in the buffer region $[0,L]$,
this non-unitarity becomes localized and, during time propagation, the wavefunctions have their norm altered 
(increased or decreased) when overlapping with that region. 
In order to absorb, the imaginary part of $V_{\rm CAP}(x)$ must take negative values since
the presence of a negative imaginary potential at the exponent of the propagation operator of \eqref{eq:ucap}
induces an exponential damping of wavepackets. 

Among the many possible choices, the monomial CAP represents one of the most discussed in the  
literature~\cite{Riss:1996iu,Zavin:1998dt,MUGA:2004jy,Jhala:2010cr}. 
It consists of a purely imaginary potential of the form
\begin{equation}\label{eq:capmono}
	V_{\rm CAP}(x)=V_{\rm m}(x)=  \left\{ 
	\begin{array}{ll}
		0 & \mbox{if $x < 0$} \\
		i\alpha x^{b} & \mbox{if $0\leq x \leq L$} 
	\end{array}
	\right. \,. 
\end{equation}

\begin{figure}[ht]
	\includegraphics[width=\columnwidth]{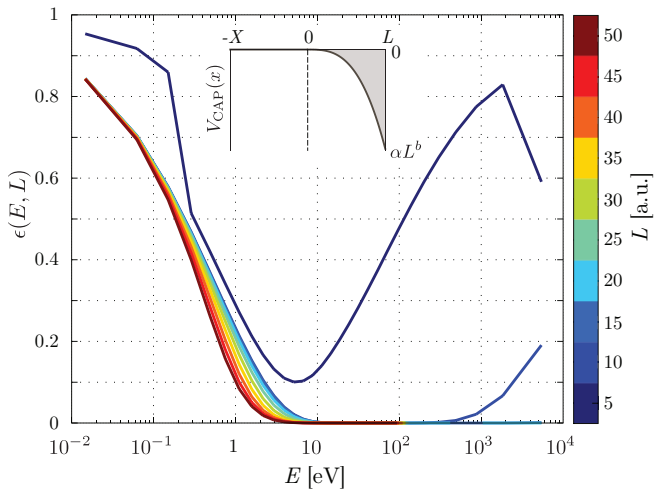} 
	\caption{Reflection error $\epsilon(E,L)$ as a function of energy $E$ and boundary width $L$ for a monomial 
  complex potential with $b=3$ and $\alpha=-0.003$~a.u.
  The functional form of $V_{\rm CAP}$ on the imaginary axis is shown in the inset.
  \label{fig:cap_poly}} 
\end{figure}
To act as an absorber, a monomial CAP must have $\alpha<0$, while $b$ can be any number greater than one.
The strength of induced damping depends on the kinetic energy of the colliding wavepacket and on the width of the 
absorption region as shown in in Fig.~\ref{fig:cap_poly}.
In this case we select $\alpha=-0.003$~a.u. and $b=3$.

Except for $L<10$~a.u., the reflection error qualitatively follows the one previously discussed for  
the mask functions of Fig.~\ref{Mask_sin2:fig}.
For $L=5$~a.u.\ the CAP seems numerically unstable and at best capable to deliver a minimum 10\% reflection for $E\approx 5$~eV.
Compared to the MFA, the monomial CAP seems to be less sensitive to changes in $L$.
and appears to converge faster to an asymptotic shape as $L\rightarrow\infty$.
This behavior can be explained by observing that the maximum value of the potential in \eqref{eq:capmono}
scales as $i \alpha L^b$, thus introducing an additional dependence of $\epsilon(E,L)$ on $L$.
This additional dependence makes the reflection more quickly with respect to $L$
and directs the attention to the dependence of
reflection on the values of the CAP on the edge.
The relationship between CAPs and mask functions will be studied in the next section.
Generally if one lets the CAP diverge at the boundary, it will behave roughly like a mask function approaching zero.

In order to disentangle the effect of the boundary width $L$ from that of the maximum 
value of $V_{\rm CAP}$ in $[0,L]$ we study the case of a $\sin^2$ potential
\begin{equation}\label{eq:capsin2}
	V_{\rm CAP}(x)=V_{\rm s}(x)= \left\{ 
	\begin{array}{ll}
		0 & \mbox{if $x < 0$} \\
		i\eta \sin^{2}\left[ \frac{x\pi}{2L} \right] & \mbox{if $0\leq x \leq L$} 
	\end{array}
	\right. \,. 
\end{equation}
Compared with the CAP in \eqref{eq:capmono} this potential smoothly increases from zero 
to its maximum value at the edge of the box $V_{\rm s}(L)=\eta$, and, like the monomial one, it absorbs 
only for negative values of $\eta$. 
\begin{figure}[ht]
	\includegraphics[width=\columnwidth]{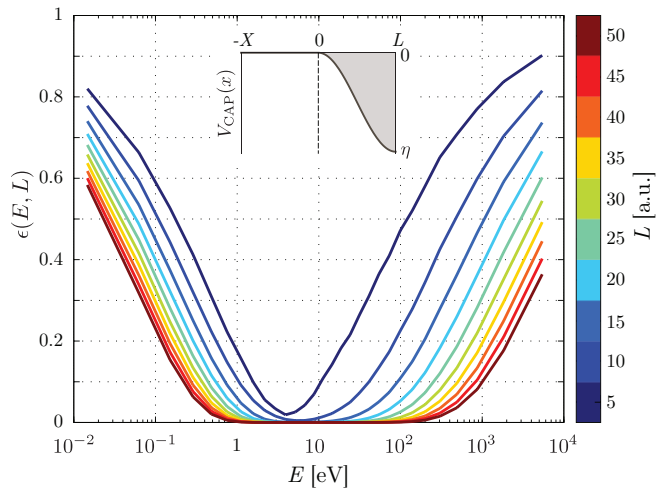} 
	\caption{Complex absorbing potential reflection error $\epsilon(E,L)$ as function of 
  energy $E$ and boundary width $L$, for a $\sin^2$ CAP as defined in \eqref{eq:capsin2}
  with $\eta=-0.2$~a.u.
  The functional shape of $V_{\rm CAP}$ on the imaginary axis is shown in the inset.
  \label{Mask_cap:fig}} 
\end{figure}

The reflection error for a $\sin^2$ CAP with $\eta=-0.2$ is shown in Fig~\ref{Mask_cap:fig}.  
Compared with the monomial CAP in Fig.~\ref{fig:cap_poly} the survival probability displays a quite 
different behavior presenting a flat minimum centered around $E=10$~eV increasingly extending with $L$. 
The low energy absorption is globally better than the monomial CAP but quickly deteriorates for high 
kinetic energies, and $\epsilon(E,L)$ is more sensitive to the changes of $\eta$.
The failure at high energies occurs when the wavepackets are fast enough to reflect at $x=L$ and exit the absorbing region
before the CAP can absorb them.

\begin{figure}[ht]
	\includegraphics[width=\columnwidth]{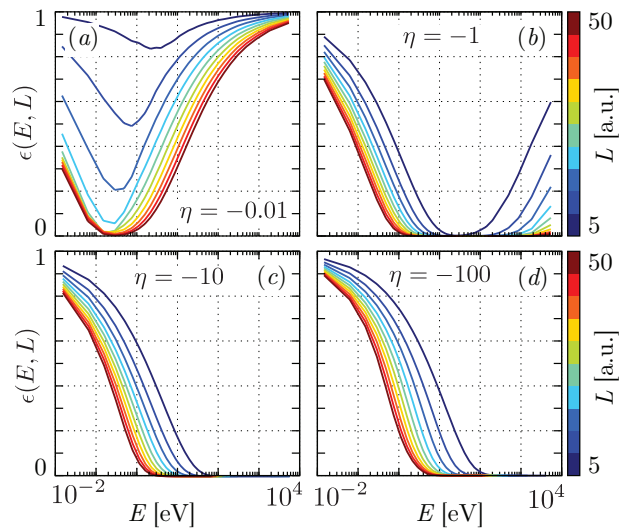} 
	\caption{Like Fig.~\ref{Mask_cap:fig}, but for different values of the CAP height $\eta$: (a) $\eta =-0.01$~a.u., 
  (b) $\eta = -1$~a.u., (c) $\eta=-10$~a.u., (d) $\eta=-100$~a.u.
  \label{Mask_cap_panel:fig}} 
\end{figure}
The reflection error is shown in Fig.~\ref{Mask_cap_panel:fig} for 
selected values of the potential height $\eta$.
For small $\eta$ the reflection error has a peaked minimum that widens up with $L$. 
As $\eta$ increases, the absorption window widens and the minimum moves to higher energies quickly 
overrunning the plot range.

We conclude the current section by presenting an example where CAPs  
can be used in time-dependent calculations.
The calculation of the continuous part of the absorption spectrum of a hydrogen atom is 
an illustrative application of CAPs to real-time propagation of the Schr\"odinger equation.
We consider the simple case of a one-dimensional hydrogen atom with a softened Coulomb interaction described by the Hamiltonian
\begin{equation}
  H_0=-\frac{1}{2}\frac{\partial^2}{\partial x^2} -\frac{1}{\sqrt{x^2+2}}\,.
\end{equation}
In the spectrum, the absorption lines appear in correspondence with allowed transitions from ground to excited states. 
They are positioned at photon energies precisely equal to the energy difference between such states, and their strength is 
proportional to the dipole matrix element connecting the states.
At photon energies above the first ionization threshold, the spectrum thus involves matrix elements connecting to continuum states.
A good description of the spectrum in this region is therefore ultimately linked to a good representation of continuum 
states in the same energy region.

Instead of directly evaluating the dipole matrix elements we here follow the time evolution of the density.
The optical absorption properties can be easily calculated, in the linear response, by analyzing the time dependence 
of the dipole moment subject to a small initial kick perturbation (see for instance chapter 7 of 
Ref.~\cite{Marques:2011ud}). 
To this end we discretize spatial and temporal coordinates with $\Delta x=0.1$~a.u., $\Delta t=0.01$~a.u., and 
perform time propagations with different boundary absorbers for a maximum time of 1000~a.u.
\begin{figure}[ht]
	\includegraphics[width=\columnwidth]{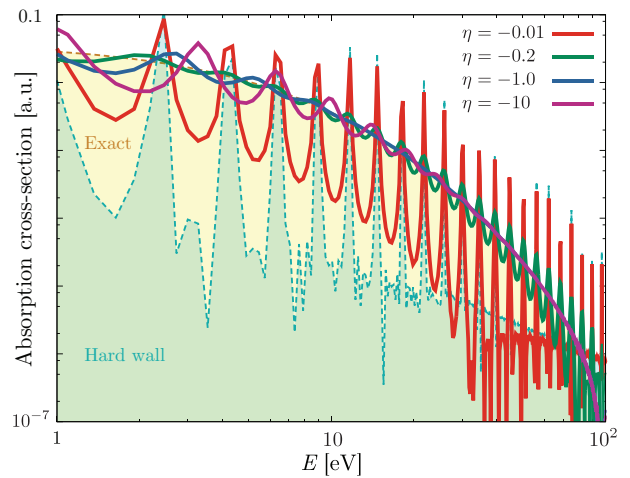} 
	\caption{One-dimensional hydrogen atom absorption cross-section (bilogarithmic scale) in the continuum. 
  The plot has been centered so that the lowest energy corresponds to the ionization threshold 13.6~eV. 
  Dashed lines over filled-color areas represent the spectrum for hard-wall boundary conditions (cyan) and a well-converged \emph{exact} solution calculated
  with a box of size 6000~a.u.\ (yellow).
  Fully drawn lines are the spectrum with a $\sin^2$ CAP defined by \eqref{eq:capsin2}, width $L=10$~a.u.\ and having 
  different heights: $\eta=-0.01$~a.u.\ (red), $\eta=-0.2$~a.u.\ (green), $\eta=-1$~a.u.\ (blue), 
  and $\eta=-10$~a.u.\ (purple).  
  \label{fig:habs}} 
\end{figure}

First we analyze the situation where no CAP is present. 
To this end we perform the time evolution in a 60~a.u. box with hard wall boundary conditions. 
The resulting spectrum centered in the continuum region (i.e. for photon energies above the ionization threshold) 
is shown in Fig.~\ref{fig:habs} (cyan dashed line over pale green area).
It is highly structured and presents a series of peaks where a smooth behavior should be expected.
These peaks can be associated with transitions to unphysical box-states, i.e.\ the eigenstates associated with the 
infinite square well embedding the system.
The continuum, constituted by such discrete states, is thus strongly dependent on the box size and far from being  
a good approximation for the real one.

In order to avoid reflection, or equivalently to increase the number of box levels per unit of energy, a much larger 
simulation box has to be employed.
In the current case, a box of 6000~a.u.\ is enough to avoid reflections and completely contain the wavefunction at the 
end of the time propagation.
In contrast to the previous case, here the spectrum (yellow dashed line over pale yellow area) presents a smooth behavior characteristic 
of the continuum region.  
The goal of a CAP would be to reproduce the same spectrum with a much smaller simulation box.

To this end we fix the box size to 60~a.u.\ and recalculate the spectrum by placing two $L=10$ a.u.\ wide 
$\sin^2$ CAPs at the edges and vary $\eta$. 
For $\eta=-0.01$~a.u.\ the spectrum presents a marked oscillating structure with peaks located at box-state 
transition energies. 
The effect of this CAP is here limited to a smearing between peaks plus a small increase of the absorption. 
For $\eta=-0.2$~a.u.\ the spectrum becomes smoother for low energies. 
In this region it lies on top of the exact solution until it departs from it as the energy becomes larger than 10~eV.
A further decrease of $\eta$ moves the smooth region forward to higher energies leaving oscillations in the lower 
energy end.

None of these CAPs is good enough to reproduce the exact results for the whole range of energies but only in a 
limited region that changes with $\eta$.
This behavior corresponds well to the one we observed for the reflection error $\epsilon(E,L)$ 
in Fig.~\ref{Mask_cap:fig} and Fig.~\ref{Mask_cap_panel:fig}.
We therefore conclude that $\epsilon(E,L)$, although being an approximation to the survival probability $S(E,L)$,
can provide useful information to assess and control reflections in practical calculations.

\section{Connecting MFAs and CAPs} 
\label{sec:mask_and_cap}
This section is devoted to the discussion of the mathematical connections $a$ and $b$ in the scheme of Fig.~\ref{fig:scheme}.
The existence of such a connection has been known in the literature since long time ago~\cite{Kosloff:1986et,Chelkowski:1998ec} but, to 
the best of the authors' knowledge, has never been presented in a unified framework.

We begin with the problem of linking an MFA to a CAP and therefore with the problem 
of casting the first into the formalism of the second.
The solution can be simply obtained by comparing the time propagation operators in the two formalisms. 
Starting from \eqref{eq:MaskProp} we can define an infinitesimal mask function time propagation operator as
\begin{eqnarray}
	\label{eq:umask} 
  U_{M}(t+\Delta t, t)&=&M(x) U(t+\Delta t, t) = \nonumber\\
	&=&M(x) e^{-i H_0 \Delta t} = \nonumber \\
	&=& e^{\tilde{M}(x)}e^{-i H_0 \Delta t}\,, 
\end{eqnarray}
where $H_0$ is the physical Hamiltonian and 
\begin{equation}\label{eq:logmask}
  \tilde{M}(x)=\ln[M(x) ]  
\end{equation}
is a logarithmic mask function.
By requiring the two propagators in \eqref{eq:umask} and \eqref{eq:ucap} to be equal
it is possible to derive an explicit form for the CAP $V^M_{\rm CAP}(x)$ associated with the mask function $M(x)$.
The resulting equation can be expanded using the Baker--Campbell--Hausdorff formula for exponential 
operators and, to first order in $\Delta t$, this CAP can be written as (see Appendix~\ref{sec:mask_and_cap_mapping}),
\begin{eqnarray}
	\label{eq:capmask0} 
	V^{M(1)}_{\rm CAP}(x)&= & \frac{i}{\Delta t} \tilde{M}  \\
  &+&\frac{1}{2}[\tilde{M}(x),H_0]+ \frac{1}{12}[\tilde{M}(x),[\tilde{M}(x),H_0]] \nonumber \\
	&-& \frac{1}{720}[[[[H_0,\tilde{M}(x)],\tilde{M}(x)],\tilde{M}](x),\tilde{M}(x)]+ \dots\nonumber \,.
\end{eqnarray}
This can be simplified by evaluating the commutator $[H_0,\tilde{M}(x)]$.
The result is a closed formula for the mask function--CAP approximation to first order in $\Delta t$:
\begin{align}
	V^{M(1)}_{\rm CAP}(x)&= \frac{i}{\Delta t} \tilde{M}(x)
	- \frac{
	\dee\tilde{M}(x)}{
	\dee x}\frac{
	\partial}{
	\partial x}
        \nonumber\\&\quad
        +\frac{1}{6}\left(\frac{
	\dee\tilde M(x)}{
	\dee x}\right)^{2} - \frac{1}{2}\frac{
	\dee^{2}\tilde{M}(x)}{
	\dee x^{2}}\,.\label{eq:capmask} 
\end{align}
The mask function derived CAP therefore contains the differential operator $\partial/\partial x$  and is not purely imaginary, but has 
components also on the real axis.
CAPs that are not strictly imaginary have been studied in the literature, and it has been pointed out that the inclusion of a real component
in the potential may lead to enhanced absorption~\cite{MUGA:2004jy}.

For small values of $\Delta t$ the CAP in \eqref{eq:capmask} is dominated by the first 
term $i\tilde{M}(x)/\Delta t $ which is purely imaginary. 
For every mask function $0\leq M(x)\leq 1$ this term is on the negative imaginary axis, and so $V^{M(1)}_{\rm CAP}(x)$
acts as an absorber.

\begin{figure}[ht]
	\includegraphics[width=\columnwidth]{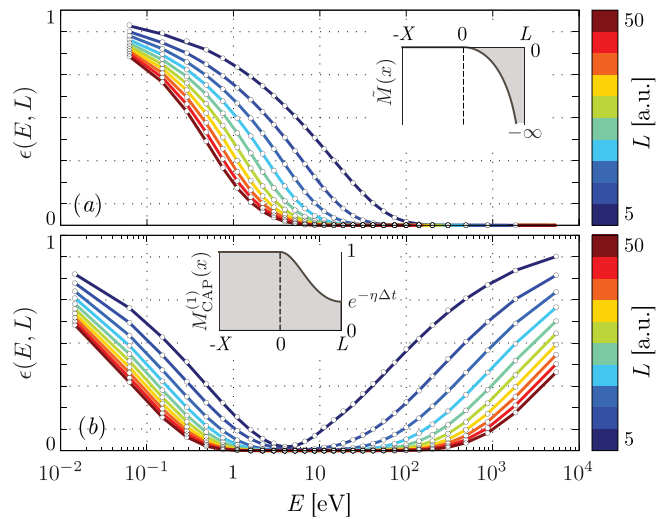} 
	\caption{Comparison of reflection errors $\epsilon(E,L)$ for the same absorber expressed 
  as CAP or mask function. 
  Panel (a) the mask-function absorber of \eqref{eq:Mask_sin2} (points) compared to its
  representation as CAP \eqref{eq:capmask} (solid lines). The inset depicts $\tilde{M}(x)$. 
  Panel (b) $\sin^2$ CAP as of \eqref{eq:capsin2} with $\eta=0.2$~a.u.\ (points) expressed as a mask function through
  \eqref{eq:maskcap} (solid lines). In the inset the first-order mask function $M_{\rm CAP}^{(1)}(x)$
  is pictured.  
  \label{fig:maskcap_capmask}} 
\end{figure}
In Fig.~\ref{fig:maskcap_capmask}~(a) we present a comparison of the reflection error $\epsilon(E,L)$
calculated with the MFAs defined in \eqref{eq:Mask_sin2} (points) and the one 
obtained using the CAP of \eqref{eq:capmask} (solid lines). 
The agreement between the curves for different $L$ underlines the good quality 
of the first-order approximation for the time step used.
In this case $M(L)=0$, and thus $\tilde{M}(x) \rightarrow -\infty$ as $x\rightarrow L$.

We now turn to the derivation of the inverse relation, i.e. the problem of expressing a CAP as a mask function.
Once again, the correspondence is derived by requiring the evolution operators to be equal. 
Using Zassenhaus' expansion for $U_{\rm CAP}(t+\Delta t, t)$ we have
\begin{align}\label{eq:maskcap}
		M_{\rm CAP}(x)&=e^{-i  V_{\rm CAP}(x) \Delta t}  \nonumber\\
  &\quad\times e^{\left[ 2 \frac{
  	\dee V_{\rm CAP}(x)}{
  	\dee x } \frac{
  	\partial }{
  	\partial x } + \frac{
  	\dee^2 V_{\rm CAP}(x)}{
  	\dee x^2 } \right]\frac{\Delta t^2 }{2} } \times \cdots
\end{align}
Similar to the previous case, the CAP mask function is here complex-valued.
This does, however, not affect the splitting property of  
\eqref{eq:masksplit} which also includes this situation.
When $V_{\rm CAP}(x)$ is a purely imaginary potential, the first-order term of \eqref{eq:maskcap},
\begin{equation}\label{eq:maskcap1}
 M_{\rm CAP}^{(1)}(x)=e^{-i \Delta t V_{\rm CAP}(x)} \,,  
\end{equation}
becomes a real function.

A numerical comparison for a $\sin^2$ CAP ($\eta=0.2$~a.u.) expressed in the two formalisms is shown on
Fig.~\ref{fig:maskcap_capmask}~(b).
It shows that \eqref{eq:maskcap1} yields two numerically identical absorbing properties and that,  
for practical applications, the two methods are equivalent.
Further, the results indicate that the value of a mask function
at the edge of the box is connected with the position of the minimum in $\epsilon(E,L)$.    
This can be easily understood by comparison with $V_{\rm CAP}(L)$.

Finally, we observe that compared with the CAPs whose values may range from 0 to $-\infty$ in $[0,L]$, 
the mask function spans a much more limited range $[0,1]$. 
As we will discuss in the next section, this fact has positive numerical implications possibly leading to stabler propagation schemes.


\section{Smooth-exterior scaling (SES)} 

\label{sec:ses} 
We conclude our inventory of ABs by discussing the effect of a smooth-exterior
scaling of the Hamiltonian as a reflection absorber.
This approach is based on a generalization of the original, uniform complex scaling 
method~\cite{aguilarcombes,Balslev:1971ez}. 
In SES the spatial coordinates of the Hamiltonian are rotated into the complex plane by an angle $\theta$ outside
a certain region by means of the transformation~\cite{Simon:1979tc,Moiseyev:1988ea,Moiseyev:2011tx}: $x\rightarrow F(x)$.
This transformation has to be such that it asymptotically recovers the linear scaling transformation 
\begin{equation}\label{eq:fxlim}
  F(x) \rightarrow x e^{i\theta}\quad {\rm for}\quad x\rightarrow\infty\,.
\end{equation}
As can be easily shown, given a function $g(x)$ smoothly increasing from 0 to 1 close to the onset 
of the boundary region $x=0$, the family of transformations 
\begin{equation}\label{eq:sesf}
  f(x)=\frac{\dee F(x)}{\dee x} = 1 + (e^{i\theta} -1)g(x)
\end{equation}
satisfies the limit condition of \eqref{eq:fxlim}.
For every choice of $g(x)$, it is then possible to recover $F(x)$ by direct 
integration of \eqref{eq:sesf}.

Following a SES transformation, the new Hamiltonian turns non-Hermitian and, like for CAPs, 
the time-evolution operator becomes non-unitary.
It is further possible to push the connection with CAPs by defining a $V_{\rm CAP}^{\rm S}(x)$ that added to the Hamiltonian
$H_0$ implements the SES transformation~\cite{Moiseyev:1999hs}.
The SES CAP is
\begin{equation}\label{eq:vses}
  V_{\rm CAP}^{\rm S}(x)= V_0(x) + V_1(x)\frac{\partial }{\partial x} + V_2(x)\frac{\partial^2 }{\partial x^2} 
\end{equation} 
where
\begin{align}
  V_0(x)&=\frac{1}{4 f^3(x)}\frac{\dee^2 f(x)}{\dee x^2}-\frac{5}{8 f^4(x)}\left(\frac{\dee f(x)}{\dee x}\right)^2  \\
  V_1(x)&=\frac{1}{f^3(x)}\frac{\dee f(x)}{\dee x}\\
  V_2(x)&=\frac{1}{2}\left(1-\frac{1}{f^2(x)}\right)
\end{align}
derive from an evaluation of the kinetic operator on $F(x)$ together with a scaling of the volume element 
${\rm d}z =f(x) {\rm d}x$. 

The CAP defined in \eqref{eq:vses} contains the differential operators $\partial/\partial x$ and $\partial^2/\partial x^2$
that make it actively change its shape as a wavepacket moves in.
It constitute one of the simplest way to implement SES in real-space codes
and explicitly illustrates the existence of the connection $c$ in Fig~\ref{fig:scheme}.
The existence of the inverse connection $d$ has been discussed in the literature~\cite{Riss:1993kc}
indicating the possibility to map a given CAP with an SES transformation~\cite{Santra:2006kd}.
It is worth to note that such connection is limited to the absorption of outgoing wavepackets only, and that 
ECS provides more than just a simple absorber.
In fact, a time propagation correctly accounting for the evolution of left and right eigenstates of the 
exterior complex scaled Hamiltonian would offer, in principle, the possibility to recover the electron dynamics in 
the outer region and allow incoming wavepackets to reenter the inner region~\cite{McCurdy:1991jp, Scrinzi:2010jv}.

A popular choice for the function $g(x)$ in \eqref{eq:sesf} is provided by 
\begin{equation}\label{eq:sesgtanh}
  g(x)= \frac{1}{2}(1+\tanh[\lambda x])\,.
\end{equation}
Integrating over $x$ in \eqref{eq:sesf} then yields
\begin{equation}\label{eq:sesftanh}
  F(x)=x+(e^{i\theta}-1)\left(\frac{x}{2} +\frac{1}{2\lambda}\log[\cosh[\lambda x]] \right)\,.
\end{equation}
Similar SES has been employed in static calculations of resonance lifetimes~\cite{Sajeev:2006vf} 
as well as boundary absorber in real-time propagations~\cite{Zavin:1998dt,Shemer:2005jh, Kalita:2011do}.

\begin{figure}[ht]
	\includegraphics[width=\columnwidth]{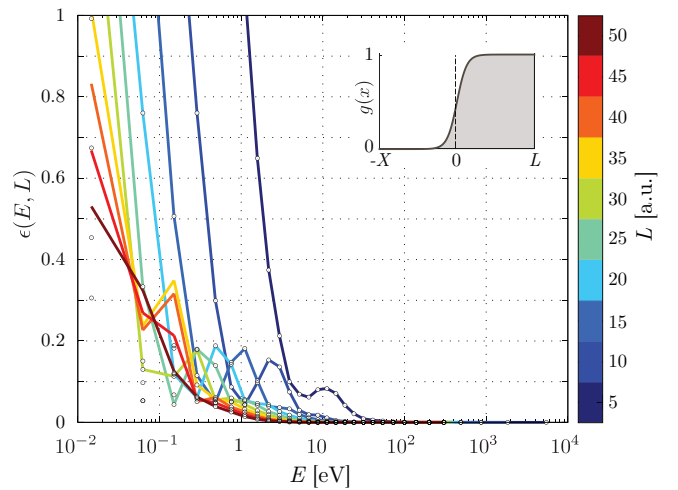} 
	\caption{Reflection error $\epsilon(E,L)$ (solid lines) for an SES CAP with $g(x)$ (inset) 
  defined by \eqref{eq:sesgtanh} with scaling angle $\theta = 0.2$ and smoothing parameter $\lambda=1$. 
  The same absorber but expressed in the mask function formalism of \eqref{eq:maskcap1} is plotted with points. 
  \label{fig:sestanh}} 
\end{figure}

In Fig.~\ref{fig:sestanh} (solid lines) we show the absorption properties of $V_{\rm CAP}^{\rm S}(x)$ associated 
with a $g(x)$ defined by \eqref{eq:sesgtanh} having $\lambda=1$ and $\theta = 0.2$.
The reflection error at any given $L$ presents similar traits quickly decaying for large 
energies.
As the energy is lowered, $\epsilon(E,L)$ steeply increases and exceeds 1 indicating the presence of an instability 
region where the SES amplifies the norm instead of reducing it.
This region is pushed towards lower energies as $L$ increases;
for $L>30$~a.u. we observe numerical instability only for energies $E<0.063$~eV.

The scaling transformation \eqref{eq:sesftanh} implements a smooth transition of the wavefunction from $\psi(x)$ for 
$x\ll0$ to $e^{i \theta/2}\psi(F(x))$ for $x\gg0$ (the factor $e^{i \theta/2}$ is necessary to ensure the 
unitarity of the scaling transformation~\cite{Scrinzi:2010jv}) on a region with spatial extension 
proportional to $1/\lambda$.
Spurious reflections are introduced when the grid is not fine enough to capture the phase jump of $e^{i \theta/2}$ 
in the transition region.
This can be identified with the appearance of a bump structure in the reflection curves.

The connections between ABs built so far, Fig.~\ref{fig:scheme}, allows us to proceed a step further in the chain 
and recast SES as a MFA through the composition of $b$ and $c$ links.
The points in Fig.~\ref{fig:sestanh} represent the reflection error for an MFA associated to 
$V_{\rm CAP}^{\rm S}(x)$ through \eqref{eq:maskcap1}.
The behavior here is slightly different from what we observed in the previous section; 
for $E > 0.15$~eV, $\epsilon(E,L)$ closely reproduces the results obtained with $V_{\rm CAP}^{\rm S}(x)$,
while for lower energies it absorbs better.
This indicates a better numerical stability of the mask function formalism compared to the SES CAP one,
at least in the present implementation.

Since the SES transformation \eqref{eq:sesftanh} extends beyond the boundary region $[0,L]$ 
into $[-X,0]$ a direct comparison of the reflection curves in Fig.~\ref{fig:sestanh} with the ones presented in 
the previous sections is unfair.
In order to enable the comparison we introduce a new function 
\begin{equation}\label{eq:sesgsin2}
	g(x)= \left\{ 
	\begin{array}{ll}
		0 & \mbox{if $x < 0$} \\
		\sin^{2}\left[ \frac{x\pi}{2L} \right] & \mbox{if $0\leq x \leq L$}\\
		1 & \mbox{if $x > L$}      
	\end{array}
	\right. \,, 
\end{equation}
localized in the boundary region.
The new transformation is thus turning into the complex plane only for $x>0$ reaching the asymptotic condition 
$\approx xe^{i \theta}$ at the border of the boundary region $[0,L]$:
\begin{equation}\label{eq:sesfsin2}
	F(x)= \left\{ 
	\begin{array}{ll}
		x & \mbox{if $x < 0$} \\
		 x+(e^{i \theta} -1)\left(\frac{x}{2}-\frac{L }{2 \pi }\sin\left[\frac{\pi  x}{L}\right]\right)  & \mbox{if $0\leq x \leq L$}\\
		(x-\frac{L}{2}) e^{i \theta}+\frac{L}{2} & \mbox{if $x > L$}      
	\end{array}
	\right. . 
\end{equation}

\begin{figure}[ht]
  \includegraphics[width=\columnwidth]{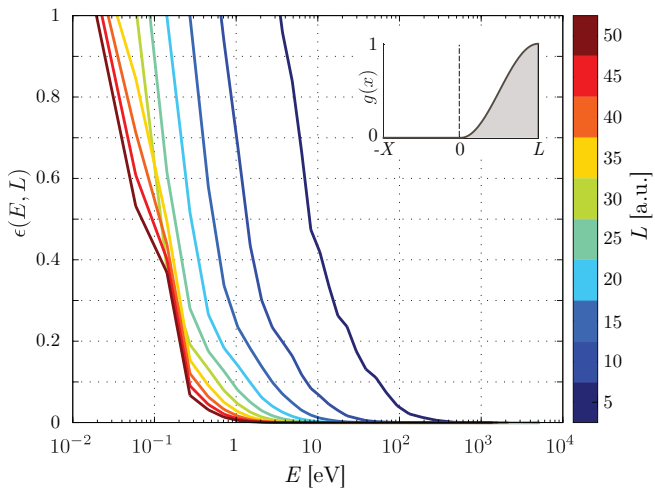} 
	\caption{Reflection error $\epsilon(E,L)$ (solid lines) for an SES CAP with $g(x)$ (inset) 
  defined by \eqref{eq:sesgsin2} with scaling angle $\theta = 0.2$. 
  \label{fig:capsessin2}} 
\end{figure}

Compared with the scaling transformation \eqref{eq:sesftanh} the new one \eqref{eq:sesfsin2} implements 
the smooth transition on the whole region $[0,L]$, and thus offer more flexibility to handle the phase jump.
As a result, the reflection error shown in Fig.~\ref{fig:capsessin2} for $\theta = 0.2$,
is qualitatively different from the one previously observed and does not present any bump.
Obviously, owing to the smaller effective region, it presents higher reflection levels compared with 
Fig.~\ref{fig:sestanh}. 
Numerical instabilities are still present and, for $L>30$ a.u., appear at $E<0.2$~eV.

From a direct comparison it is apparent that, the reflection properties offered by SES present 
the largest absorption region observed among all the ABs discussed in the present paper 
and thus strengthen the belief of SES being the key to 
reflection-free CAPs.

\section{Conclusions} 

\label{sec:conclusions}

In this paper we studied the problem of modeling the dynamics of extended
states in finite volumes with the use of absorbing boundaries.

We derived a method that permits the evaluation of the absorption properties of any absorber based 
on the numerical time propagations of localized wavepackets.
This method served to systematically illustrate the main features of the three most important boundary families
known in the literature:  
mask function absorbers, complex absorbing potentials, and smooth exterior complex scaling potentials.

We showed how the calculated numerical reflection properties can be directly employed to assess the quality of the  
absorption spectrum of the one-dimensional hydrogen atom in the continuum. 
This indicates the possibility to employ the reflection curves presented in a large number of 
situations involving real atoms and molecules. 

Finally we discussed the intimate connection between the three families of ABs.
Although being piecewise discussed in the literature, in this work we illustrated this network of connections
in a unified picture.
This was done by pairing analytical derivations with numerical examples. 
Furthermore, we showed how in certain situations, mask function boundaries can provide a stabler formalism for 
numerical applications. 

In the recent years lot of attention as been posed on the description of dynamical processes involving 
continuum states in different sectors of physics ranging from ultra-fast, ultra-intense laser physics to 
molecular transport. 
From the theoretical standpoint this poses tremendous challenges that need to be overcome.
The consolidation of a common ground for the illustration and classification of ABs presented in this work
is an important step towards the development of numerical tools capable to accurately capture these complex phenomena.



\section*{ACKNOWLEDGMENTS} 
\label{sec:acknowledgments}
The authors acknowledge financial support from the European Research Council Advanced Grant DYNamo
(ERC-2010-AdG-267374), the European Commission project CRONOS (Grant number 280879-2 CRONOS CP-FP7), 
Spanish Grants (FIS2010-21282-C02-01 and PIB2010US-00652), and 
Grupos Consolidados UPV/EHU del Gobierno Vasco (IT-578-13).


\appendix 

\section{MFA and CAP } 
\label{sec:mask_and_cap_mapping}

The connection between CAPs and mask functions is established using the Baker--Campbell--Hausdorff 
formula which expresses the product of the exponentials of two operators $X$ and $Y$ as a single exponential
\begin{align}
	e^{X}e^{Y}=e^Z
\end{align}
with
\begin{align}
	Z&=X+Y+\frac{1}{2}[X,Y]+\frac{1}{12}[X,[X,Y]] \\
	&\quad-\frac{1}{12}[Y,[X,Y]] - \frac{1}{24}[Y,[X,[X,Y]]] + \dots \nonumber 
\end{align}
Here $[X,Y]=XY-YX$ is a commutator. In \eqref{eq:umask} the operators are $X=\tilde{M}$ and $Y=-i H_0 \Delta t$. 
Substituting and ordering by powers of $\Delta t$,
\begin{align}
	Z&= -i\Delta t \left(\frac{i}{\Delta t} \tilde{M}+H_0+\frac{1}{2}[\tilde{M},H_0]\right. \nonumber\\
	&\quad+\left.\frac{1}{12}[\tilde{M},[\tilde{M},H_0]]  - \frac{1}{720}[[[[H_0,\tilde{M}],\tilde{M}],\tilde{M}],\tilde{M}]+ \dots\right) \nonumber\\
	&\quad-\Delta t^2 \left( - \frac{1}{12}[H_0,[\tilde{M},H_0]] + \dots\right) + O(\Delta t^3) 
\end{align}
we obtain an explicit form for the exponent of the finite-difference time propagator.
The formula in \eqref{eq:capmask0} can be easily obtained by comparison with \eqref{eq:umask} to 
first order in $\Delta t$.

The results can be further simplified by direct evaluation of the commutators. 
Even in the presence of an external potential in $H_0$, only the kinetic operator yields
a contribution $[-\nabla^2 /2,\tilde{M}(x)]$ to the commutator. 
In one dimension, nested commutators containing a single instance of $H_0$ and more than two $\tilde{M}(x)$ are zero, and \eqref{eq:capmask0} can be written in a closed form.




%


\end{document}